%% file: systemMatrixPaperSpringer.tex
\RequirePackage{fix-cm}
\documentclass[smallcondensed]{svjour3}     
\smartqed  
\usepackage{graphicx}
\graphicspath{{Images/}} 
\usepackage{mathptmx}      
\usepackage[utf8]{inputenc} 
\usepackage[OT1]{fontenc}
\usepackage[english]{babel}
\usepackage{amsmath,amssymb,euscript}
\usepackage{pgfplots}
\pgfplotsset{compat=newest}
\pgfplotsset{plot coordinates/math parser=false}
\newlength\figureheight
\newlength\figurewidth 
\usepackage{subfig}
\usepackage{orcidlink}

\usepackage{comment}
\newcommand{\pd}[2]{\frac{\partial #1}{\partial #2}}
\newcommand{\dr}{\,\text{d} r}

\newcommand{\ddt}{\frac{\text{d}}{\text{d}t} }
\newcommand{\correct}[1]{\textcolor{black}{#1}}
 \journalname{Sensing and Imaging}
\begin{document}

\title{Modeling Magnetic Particle Imaging for Dynamic Tracer Distributions}



\author{Christina Brandt  \and
        Christiane Schmidt  
}
%

\institute{Christina Brandt\orcidlink{0000-0002-8368-6807}
	\and
	Christiane Schmidt\orcidlink{0000-0003-4180-0971} \at
	Universität Hamburg, 
	Department of Mathematics, 
	Bundesstraße 55, 
	20146 Hamburg, 
	Germany\\ 
	\email{christiane.schmidt@uni-hamburg.de}
}

\date{Received: date / Accepted: date}

\maketitle

\begin{abstract}
\input{Abstract}
\end{abstract}

\section{Introduction}
	\label{intro}
	\input{Intro}
\section{Basic principles of magnetic particle imaging}
	\label{sec:MPIbasics}
	\input{MPIbasics}
\section{\correct{Dynamic forward model}}
	\label{sec:newModel}
	\input{MPImodelNew}
\section{Relevance of the dynamic model}
	\label{sec:relevance}
	\input{Relevance}
\section{Comparing reconstructions with the dynamic and static model}
	\label{sec:examples}
	\input{Recos}
\section{Discussion and conclusion}
	\label{sec:conclusion}
	\input{Conclusion}
%
%

%
%
\section*{Declarations}
\paragraph{Funding} No funds, grants, or other support was received.
\paragraph{Conflict of interest} The authors have no conflicts of interest to declare that are relevant to the content of this article.
\paragraph{Availability of Data} The data sets generated during the experiments are available from the corresponding author on reasonable request.
\paragraph{Code availability} The code used for the simulations is available from the corresponding author on reasonable request.
%
%
\bibliographystyle{spmpsci}      
\bibliography{systemMatrixPaperLitVZ}   
%
%
%
%
\end{document}

%% file: Abstract.tex
%
%
%
Magnetic Particle Imaging (MPI) is a promising tracer-based, functional medical imaging technique which measures the non-linear \correct{magnetization response of magnetic nanoparticles} to a dynamic magnetic field. For image reconstruction, system matrices from time-consuming calibration scans are used predominantly. Finding modeled forward operators for magnetic particle imaging, which are able to compete with measured matrices in practice, is an ongoing topic of research.
The existing models for magnetic particle imaging are by design not suitable for \correct{arbitrary} dynamic tracer concentrations. Neither modeled nor measured system matrices account for changes in the concentration during a single scanning cycle.\\
In this paper we present a new MPI forward model for dynamic concentrations. A \correct{static} model will be introduced briefly, followed by the changes due to the dynamic behavior of the tracer concentration. Furthermore, the relevance of this new extended model is examined by investigating the influence of the extension and example reconstructions with the new and the standard model.\\

%% file: Intro.tex
%
%
Magnetic Particle Imaging (MPI) is a relatively new medical imaging modality invented by Weizenecker and Gleich in 2005 \cite{gleich2005tomographic}. \correct{In this tomographic imaging technique, the non-linear magnetization response of the superparamagnetic tracer material to an external magnetic field induces a potential in the receive coils of the scanner. The spatial distribution of the magnetic particles is reconstructed from these measurements. MPI} allows for a rapid data acquisition with high temporal resolution which makes it a promising imaging device for different imaging applications, see \cite{Knopp2017} for an overview. 
%
In many of these applications, visualization of tracer dynamics is highly relevant, such as physiological diagnosis like stroke detection \cite{ludewig2017magnetic}, visualization of blood flow \cite{weizenecker2009three} or \correct{localization of medical instruments} in vascular interventions \cite{haegele2012magnetic}.\\
%
The MPI forward operator can be described by model- or measurement-based approaches \cite{gruttner2013formulation}. In a measurement-based approach the forward operator is represented by a calibration scan \cite{rahmer2012analysis,weizenecker2007simulation}. Therefore, the signal generated by a delta sample of tracer material is measured for a finite number of spatial positions.
The modeling approach describes the measurement process with physical laws \cite{knopp2009model}. Unfortunately, models usually idealize the physical setting to limit the complexity of the model. These simplifications can lead to large modeling errors and give reasons for the time consuming measurement approach being still dominant in practice.\\
%
\correct{The state of the art model} underlies the assumption of a (nearly) static concentration during the signal acquisition. \correct{This assumption is not always fulfilled. MPI is able to visualize the distribution of a liquid tracer. It can accumulate, dissipate or move e.g. with the blood flow. The behavior of the particle concentration is not static in these cases. Also time-series measurements imply a dynamic tracer distribution such that the static model is only true for piecewise constant concentrations.} The same problem is valid for the measurement approach. Since the delta sample is static during each cycle of the calibration scan, the measured system matrix does not cover dynamic behavior. 
%
\correct{Currently, the only way to reconstruct non-periodic dynamic concentrations is to reconstruct a time-series of images under the assumption of static behavior during the scan \cite{storath2017,haegele2012magnetic}. Reconstruction of periodic dynamics in magnetic particle imaging is investigated in \cite{gdaniec2020rotation} in order to reduce of artifacts induced by cardiac- or respiratory motion in multi-patch MPI. The authors use the measurement-based approach and assume limits on the velocity and periodicity of the motion. By rearranging measurements from the same motion phase into virtual frames, dynamic tracer distributions can be reconstructed by static reconstructions from the virtual frames.}\\
%
The model-based approach gives rise to various directions of research covering all components of the signal generation chain and analyses of the models. One of these directions is modeling of the magnetic behavior of \correct{magnetic nanoparticles} which was studied by Kluth \cite{kluth2018mathematical,Kluth2020IWMPI} and Weizenecker \cite{weizenecker2018fokker}. The most frequently used magnetization model is the Langevin- or equilibrium model, which is also the basis for the derivations in the following articles. \correct{The equilibrium model does not respect magnetic relaxation effects. In \cite{kluth2018mathematical}, the model is extended for different kinds of relaxation. The author presents forward models incorporating either Brownian rotation or N\'{e}el relaxation in the cases of mono- and polydisperse tracers under the assumption of single domain particles with uniaxial anisotropy.}\\
%
\correct{Based on the equilibrium model the authors of \cite{marz2016model}} derive analytical reconstruction formulae as well as numerical reconstruction schemes for two- and three-dimensional MPI. They examine and compare the ill-posedness of the reconstruction problem for different dimensions. 
A mathematical analysis of the 1D model is provided by Erb, Weinmann et al. \cite{erb2018mathematical}. They investigate properties like the ill-posedness and discover an exponential singular value decay of the reconstruction problem. 
Goodwill and Conolly \cite{goodwill2010x} follow the X-space approach. They consider the dependence of the spatial position of the field free point (FFP), which is the time-dependent volume of vanishing magnetic field strength, and the drive field of the scanner. As a result the forward operator of 1D MPI can be identified as a convolution operator. The authors extend their approach to multiple dimensions in \cite{goodwill2011multidimensional}. 
The more analytically focused article by Maass and Mertins deals with closed-form expressions for the Fourier transform of the system function for  multiple dimensions \cite{maass2019b}. The system function is related to tensor products of Chebychev polynomials of the second kind and tensor products of Bessel functions. This result might allow for analytical insights into the system function and more efficient reconstruction techniques in the future.\\
%
Another common model simplification is the assumption of ideal magnetic fields. In practice magnetic fields can be distorted which is influencing spatial signal encoding. In \cite{bringout2020new}, the authors use spherical harmonics to achieve more realistic representations of magnetic fields for FFP and field free line (FFL) scanners. A 3D forward model for non-ideal magnetic fields which can be reconstructed with the algebraic reconstruction technique (ART) is presented. Artifacts caused by distorted magnetic fields have also been investigated in the context of the X-space approach \cite{Yagiz2020}. 
\correct{These distortions especially affect multi-patch MPI, since distortions increase with the distance from the center. In the case of measurement-based reconstruction, compensation methods for displacement artifacts in multi-patch scans are studied in \cite{boberg2019generalized,Boberg2020IWMPI}.}\\
%
\correct{In this paper we present an extended MPI forward model for dynamic tracer distributions, in the discrete and continuous case, both in time- and Fourier domain. While the initial theoretical setup presented in Sec.\ref{sec:newModel} is identical to the one in \cite{gdaniec2020rotation}, our model is not limited to periodic motion and covers dynamic tracer distributions with high velocities. Furthermore, we provide simulation experiments to show the importance of the extension relative to the tracer dynamics and the impact on reconstruction quality compared to the static model. The presented approach is of special interest for blood flow measurements \cite{kaul2018magnetic} because the speed of the motion is part of the model and can be reconstructed simultaneously.}\\
%
The remainder of this paper starts with a brief introduction to the principles \correct{of an ideal MPI system} in Sec.\ref{sec:MPIbasics} and is followed by presenting the standard modeling approach in Sec.\ref{sec:oldModel} which we will extend to arbitrary dynamic tracer distributions in Sec.\ref{sec:newModel}. 
\correct{Based on the Langevin model, FFP scanners and Lissajous trajectories we investigate the influence of the extension to the signal for different kinds of dynamics in Sec.\ref{sec:relevance}, while we compare reconstructions of simulated dynamic measurements with the standard and the extended model in Sec.\ref{sec:examples}.} We close with a discussion of the results in Sec.\ref{sec:conclusion}.

%% file: MPIbasics.tex
\begin{figure}[htbp]
	\centering
	\subfloat[The selection field $H_s$ is a static gradient field with vanishing field strength in the center and linearly increasing field strength to the border of the field of view.\label{fig:selectionField}]{
		\includegraphics[height=.32\textwidth]{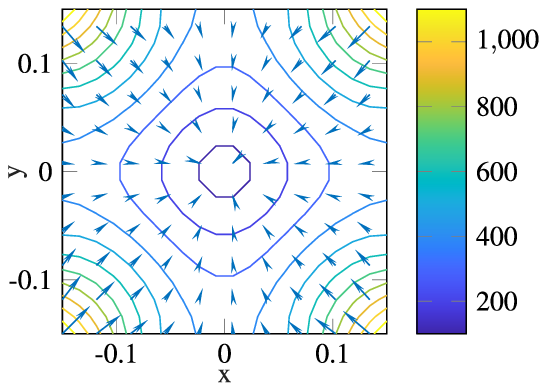}}
	\qquad
	\subfloat[The trajectory of the field free point forms a Lissajous curve during a single scan cycle.\label{fig:lissajous}]{
		\includegraphics[height=.32\textwidth]{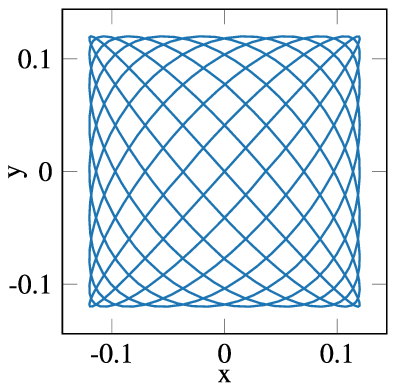}}
	\caption{Setup of the magnetic field}\label{fig:magneticFields}
\end{figure}

%
The aim of magnetic particle imaging is the  reconstruction of the multi-dimensional spatial concentration of the particles. 
Spatial encoding of the signal is realized by applying a spatially and \correct{temporally varying} magnetic field \correct{$H\in L^2(\Omega\times\mathbb{R}^+,\mathbb{R}^{3})$ where $\Omega\subset\mathbb{R}^{3}$} denotes the field of view (FOV). The magnetic field
\begin{align*}
	H(r,t)&=H_S(r)+H_D(t)
\end{align*}
consists of a spatially inhomogeneous selection field $H_S$ and a \correct{temporally varying} drive field $H_D$. The selection field is a gradient field that has a point of zero field strength in the center, the so-called field free point (FFP), and a linearly increasing field strength to the periphery (see Fig.\ref{fig:selectionField}). The drive field $H_D(t) = [a_l\sin(2\pi f_l t + \varphi_l)]_{l=1,\dots,3}$ is spatially constant but changes its magnetization over time according to a sine function in each dimension. It has three parameters per dimension, the amplitude $a_l\in \mathbb{R}$ determining the size of the field of view, the frequency $f_l\in \mathbb{R}$ defining the density of the scan trajectory and a phase shift $\varphi_l\in \mathbb{R}$ setting the starting point of the scan trajectory.
Choosing the parameters appropriately, the overlay of $H_S$ and $H_D$ forms the field $H$ which has a FFP moving through the volume of interest along a so-called Lissajous curve (see Fig.\ref{fig:lissajous}).
\\
\correct{Superparamagnetic means that the magnetic nanoparticles behave like tiny magnets, while an external magnetic field is applied. They have their own magnetic moments which are larger than their atomic moments. There is no remanent magnetization after the applied magnetic field is removed \cite{bean1959superparamagnetism}.}\\
The magnetic moment of the particles responds to \correct{temporal changes of magnetic fields}. There are different models describing the magnetic behavior of the particles which where studied in \cite{kluth2018mathematical,Kluth2020IWMPI,weizenecker2018fokker}. In Sec.\ref{sec:relevance} and \ref{sec:examples} the Langevin or equilibrium model is used but could be replaced by more complex models.
When the field free point moves over a position $r$, it causes a change in the mean magnetic moment at this location. The magnitude of the magnetization
\begin{align}
	M(r,t)&=c(r)\bar{m}(r,t),\qquad M\in H^1(\Omega\times\mathbb{R}^+),\,c \in L^2(\Omega), 
	\label{eqn:staticConcMagnetization}
\end{align}
is proportional to the tracer concentration $c:\Omega\mapsto\mathbb{R}^+$ and the mean magnetic moment \mbox{$\bar m:\Omega\times\mathbb{R}^+\mapsto\mathbb{R}^{3}$}. The Sobolev space
\begin{align*} 
	H^1(D)&=\{f \in L^2(D): D^\gamma f\in L^2(D) \text{ for } 0\leq |\gamma|\leq 1,\gamma \in \mathbb{N}^d,D\subset \mathbb{R}^d\}
\end{align*} 
denotes the space of $L^2$-functions whose \correct{first weak derivatives} are also functions in $L^2$ on the domain $D\subset \mathbb{R}^d$ with \correct{$\gamma \in \mathbb{N}^d$} being a multi-index with $|\gamma|=\sum_{i=1}^d \gamma_i$, see \cite{adams2003sobolev}.
The change in magnetization induces a current in the receive coils of the scanner. Due to the construction of the magnetic field, a measured voltage at time point $t$ can be connected to a magnetization change and thus to a certain concentration at a position $r$. 
The measurement process is described by a forward model in the following section.
\subsection{MPI forward model}
\label{sec:oldModel}
\input{MPImodelOld}

%% file: MPImodelOld.tex
%
%
\correct{The static forward model 
\begin{align}
	u (t)&=-\mu_0 \ddt \int_{\Omega}p(r)\bigl(c(r)\bar{m}(r,t)+H(r,t)\bigr)\dr
	\label{eqn:forwardModelOld}
\end{align}
describes the magetic particle imaging process in time domain \cite{knopp2012magnetic}.
The sensitivity of the receive coils $p$ is multiplied with the permeability constant $\mu_0$ and the change in magnetization which is caused by the superparamagnetic particles and the applied magnetic field $H$ from the send-coils.\\
Assuming that the signal generated by the excitation field is removed by a filter yields
\begin{align}
	u(t)&=\int_{\Omega} S(r,t)c(r)\dr,\qquad  \text{with }\; S(r,t)=-\mu_0\, p(r)\pd{}{t}\bar{m}(r,t).\label{eqn:forwardModelOldShort}
\end{align}
As defined in the previous section, the excitation field is multi-dimensional which means that the system function $S(r,t)$ maps to $\mathbb{R}^{3}$, thus $u(t) \in\mathbb{R}^{3}$ is a voltage vector, where each value is measured by the respective receive coil. 
In the discrete models in the remainder of this paper we will refer to a single component of $u(t)$ since the computations are analogous for all channels.}
%
\paragraph{Discretization}
The formulation of discrete forward models is motivated by measured system matrices and the use of numerical reconstruction methods.
Therefore, we use a basis \correct{$\{\phi_i\}_{i=1, \dots ,R} \subset L^2$} of a finite-dimensional subspace $X_R\subset L^2$. An intuitive choice are piecewise constant basis functions on equisized, pairwise disjoint quadratic or cubic domains as they are a reasonable representation of both the pixels or voxels in an image and the delta probe used for the calibration scans.\\
\correct{Using the basis functions, we obtain piecewise constant approximations of the concentration and system function
\begin{align*}
	\tilde c(r)&=\sum_{i=1}^R c_i \phi_i(r) \quad \text{and}\quad
	\tilde{S}(r,t)=\sum_{i=1}^R S_{i}(t) \phi_i(r) \quad \in X_R\,,
\end{align*}
evaluated at equidistant time sampling points $\{t_j\}_{j=1,\dots,n_T}$ with $t_j=(j-1)T_c/(n_T-1)$ and $T_c$ being the repetition time for one Lissajous cycle. Insterting $\tilde c$  in (\ref{eqn:forwardModelOldShort}) yields the following discrete forward problem 
\begin{align}
	u(t_j)&=\sum_{i=1}^R S_{i}(t_j)c_i, \quad j=1,\dots,n_T\,. \label{eqn:forwardModelDiscrete}
\end{align} 
It can also be written as a matrix vector multiplication of a concentration vector and the system matrix $\mathbf{S}$
\begin{align}
	\mathbf{u} &= \mathbf{S}\mathbf{c} \quad \text{with }\mathbf{u}\in \mathbb{C}^{n_T},\; \mathbf{S}\in \mathbb{C}^{n_T \times R},\;\mathbf{c}\in \mathbb{R}^{R}.
	\label{eqn:forwardModelOldMatrixVektor}
\end{align} 
Reconstructing the concentration vector $\mathbf{c}$ from a given measurement vector $\mathbf{u}$ is a classic inverse problem.} 
In \cite{kluth2018illposed,marz2016model}, it was shown that the multidimensional MPI reconstruction problem is severely ill-posed. Thus, computing a stable and unique solution requires regularization. Two common regularization methods in MPI are Tikhonov- and iterative regularization. The former defines a  \correct{Tikhonov functional  by adding a penalty term with a regularization parameter. The resulting minimization problem}  can then be solved by the Kaczmarz algorithm or other iterative schemes adapted to the applied regularization term. The latter option regularizes the iterative method  directly by choosing a maximum number of iterations. In both cases, the standard iterative method used in MPI is the Kaczmarz algorithm \cite{kaczmarz}.
The information from reconstructions of several channels can be combined to improve image quality.

%% file: MPImodelNew.tex
%
%
The forward model presented in the preceding section underlies the assumption of a (nearly) static concentration during the signal acquisition which might be violated in case of dynamically changing tracer distributions. MPI is able to visualize the distribution of a liquid tracer which can accumulate, dissipate or move e.g. with the blood flow. In these situations, the behavior of the particle concentration is clearly not static.\\
In practice oftentimes measured system matrices are used for MPI reconstruction. These matrices are the results of calibration scans which measure the induced voltage of a delta sample during a scanning cycle for each spatial position. This approach yields good results for static concentrations because the matrices also incorporate the transfer function of the system. Since the delta sample is static during the complete cycle the measured system matrix does not cover dynamic behavior.\\
In order to adapt the model to dynamic tracer concentrations, the magnetization function (\ref{eqn:staticConcMagnetization}) is modified such that it contains a time-dependent concentration
\begin{align}
	M(r,t)&=c(r,t)\bar{m}(r,t)\,,\qquad
	c \in H^1(\Omega\times\mathbb{R}^+),\; M,\bar{m} \in H^1(\Omega\times\mathbb{R}^+)\,.
	\label{eqn:magnetizationDyn}
\end{align}
Thus, the forward model (\ref{eqn:forwardModelOld}) changes to
\begin{align*}
	u (t)&=-\mu_0 \ddt \int_{\Omega}p(r)\bigl(c(r,t)\bar{m}(r,t)+H(r,t)\bigr)\dr.
\end{align*}
Assuming a constant coil sensitivity $p$ and that the signal generated by the excitation field is removed by a filter results in the dynamic forward model
\begin{align}
	u(t)&=\eta\ddt \int_{\Omega}c(r,t)\bar{m}(r,t)\dr, \qquad\text{with }\; \eta:=-\mu_0p\;\in \mathbb{R}\\
	&=\eta\int_{\Omega}\underbrace{\pd{\bar{m}}{t}(r,t)}_{=S_1}c(r,t)+\underbrace{\vphantom{\pd{}{t}}\bar{m}(r,t)}_{=S_2}\pd{c}{t}(r,t) \dr\label{eqn:dynModel1}\\
	&=\eta\int_{\Omega}S_1(r,t)c(r,t)+S_2(r,t)\pd{c}{t}(r,t) \dr\,, \qquad c,S_2 \in H^1(\Omega\times\mathbb{R}^+).\label{eqn:newcontinuousmodel}
\end{align}
It describes a measurement $u:\mathbb{R}^+\mapsto\mathbb{R}^{3}$ in time domain and contains a sum of two system functions $S_1,S_2:\Omega\times\mathbb{R}^+\mapsto\mathbb{R}^{3}$ multiplied with the tracer concentration $c:\Omega\times\mathbb{R}^+\mapsto\mathbb{R}^+$ and its time derivative. The derivatives $\pd{c}{t}$, $S_1=\pd{\bar{m}}{t}$ and the measurement $u$ are $L^2$-functions because the concentration $c$ and $S_2=\bar m$ are in the Sobolev space $H^1(\Omega\times\mathbb{R}^+)$. 
\paragraph{Dynamic forward model in frequency domain}
MPI measurements are usually given in frequency domain. Due to the time dependence of the concentration the static model 
\begin{align*}
	\hat u(k)&=\eta\int_{\Omega}c(r)\mathcal{F}\Bigl\{\pd{\bar{m}}{t}\Bigr\}(r,k)\dr
\end{align*}
in frequency domain changes to
\begin{align}
	\hat u(k)&=\eta\int_{\Omega}\mathcal{F}\bigl\{ c\bigr\}(r,k)\ast\mathcal{F}\Bigl\{\pd{\bar{m}}{t}\Bigr\}(r,k)
	+\mathcal{F}\Bigl\{\pd{c}{t}\Bigr\}(r,k)\ast\mathcal{F}\bigl\{\bar{m}\bigr\}(r,k) \dr\,,\\
	\label{eqn:newModelFD}
	&=\eta\int_{\Omega}\hat c(r,k)\ast\hat S_1(r,k)+\widehat{\pd{c}{t}}(r,k)\ast\hat S_2(r,k) \dr\,, \qquad \hat c,\hat S_2 \in H^1(\Omega\times\mathbb{R}^+).
\end{align}
The measurement in frequency space $\hat u:\mathbb{R}^+ \mapsto\mathbb{C}^{3}$ and the derivatives $\widehat{\pd{c}{t}}:\Omega\times\mathbb{R}^+\mapsto\mathbb{C}$, $\hat S_1:\Omega\times\mathbb{R}^+\mapsto\mathbb{C}^{3}$ are $L^2$-functions because  $\hat c\,,\hat S_2 \in H^1(\Omega\times\mathbb{R}^+)$. The convolution is only applied to the frequency components.
%
%
\paragraph{Discretization }
Using the same pixel-basis \correct{$\{\phi_i\}_{i=1, \dots ,R} \subset L^2$} for discretization as in Sec.\ref{sec:oldModel} yields the following representation of a piecewise constant dynamic concentration
\begin{align*}
	\tilde c(r,t)&=\sum_{i=1}^R c_i(t) \phi_i(r) \qquad\in X_R
\end{align*}
and analogously for the derivative $\pd{c}{t}$ and system functions $S_1$ and $S_2$. Together with the time sampling points $\{t_j\}_{j=1,\dots,n_T}$ from Sec.\ref{sec:oldModel}, we get the discretized dynamic forward problem
\begin{align}
	u(t_j)&=\eta\sum_{i=1}^R S_{1,i}(t_j)c_i(t_j)+S_{2,i}(t_j)\Bigl(\pd{c}{t}\Bigr)_i(t_j), \quad j=1,\dots,n_T\,.\label{eqn:newdiscretemodel}
\end{align}
For measurements with $F\geq1$ cycles the time sampling for the measurement and concentration changes to $\{\tau_j\}_{j=1,\dots,Fn_T}$ with $\tau_j=(j-1)FT_c/(Fn_T-1)$ while the system functions are evaluated at $t_{j\text{ mod }n_T}$. Eq.(\ref{eqn:newdiscretemodel}) is no longer a matrix-vector multiplication as in (\ref{eqn:forwardModelOldMatrixVektor}) but a sum of element-wise matrix multiplications
\begin{align*}
	\mathbf{u}&=\eta\bigl[\mathbf{S}_1\odot \mathbf{c}+ \mathbf{S}_2\odot \mathbf{D}_c\bigr]\cdot \mathbf{1}_R
\end{align*}
with $\mathbf{u}\in \mathbb{R}^{n_T},\; \mathbf{c}, \mathbf{D}_c\in \mathbb{R}^{n_T \times R},\;\mathbf{S}_1,\textbf{S}_2\in \mathbb{R}^{n_T \times R},\;\mathbf{1}_R=[1,\dots,1]^{\top}\in\mathbb{N}^{R}$ and $\odot$ being an element-wise matrix multiplication, e.g. $\mathbf{A}\odot \mathbf{B}=[a_{ij}b_{ij}]_{i=1,\dots,n}^{j=1,\dots,m}$, with matrices $\mathbf{A},\mathbf{B} \in \mathbb{R}^{n\times m}$.\\
In frequency space the same approach yields the following discrete forward problem
\begin{align*}
	\hat{\mathbf{u}}&=\eta\bigl[\hat{ \mathbf{S}}_1\ast \hat{ \mathbf{c}}+ \hat{ \mathbf{S}}_2\ast \hat{ \mathbf{D}}_c\bigr]\cdot \mathbf{1}_R
\end{align*}
with $\hat{\mathbf{u}}\in \mathbb{C}^{n_K}, \hat{\mathbf{c}}, \hat{\mathbf{D}}_c\in \mathbb{C}^{n_K \times R},\hat{ \mathbf{ S}}_1,\hat{ \mathbf{S}}_2 \in \mathbb{C}^{n_K \times R},\mathbf{1}_R=[1,\dots,1]^{\top}\in\mathbb{N}^{R}$. Again the convolution is only applied to the frequency components, i.e. the respective matrix columns. Note that a frequency domain reconstruction computes $\hat c$. To see the behavior of the concentration in time, the inverse Fourier transform needs to be applied.
Reconstruction becomes a deconvolution problem in frequency space. A typical solution approach for this ill-posed inverse problem is to make use of the convolution theorem of the Fourier transform which in this case results in time domain reconstruction.\\
The dynamic model (\ref{eqn:dynModel1}) is also mentioned in \cite{gdaniec2020rotation} but followed by strong restrictions of the dynamics such that there are no further consequences in the reconstruction process. In contrast, the models proposed in this section are valid for a broad range of dynamics, e.g. rapid changes or non-periodic behavior. The tracer distribution is required to be differentiable in time and integrable in space.

%% file: Relevance.tex
\begin{figure}[htbp]
	\centering
	\subfloat[Shape of system matrix 1 in Fourier space\label{fig:s1withcurve}]{
		\includegraphics[width=.37\textwidth]{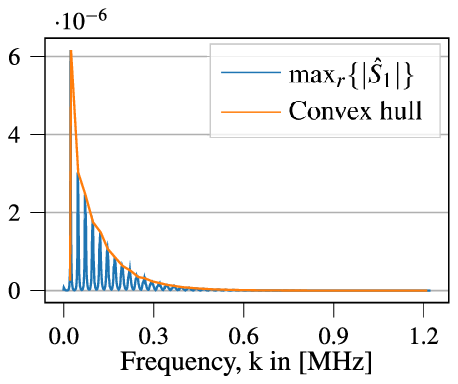}}
	\qquad
	\subfloat[Shape of system matrix 2 in Fourier space\label{fig:s2withcurve}]{
		\includegraphics[width=.37\textwidth]{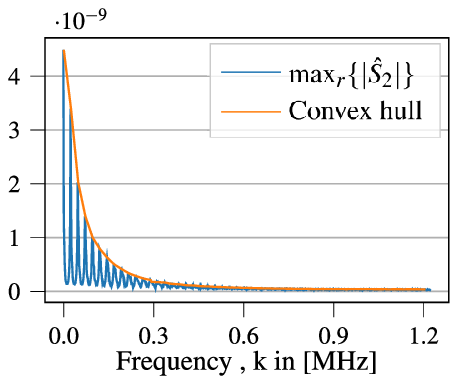}}
	\caption{Comparing the influence of the two system matrices by analyzing the  maximum over all voxels of the absolute values of the matrices in Fourier space. The orange line is an approximation of the convex hull.}\label{fig:SMswithcurves}
\end{figure}

%
%
As mentioned in Sec.\ref{sec:oldModel}, the concentration is usually assumed to be constant. The time derivative of the concentration would therefore be nearly zero and the second summand of the extended model (\ref{eqn:newcontinuousmodel}) would thus be small such that it can be neglected.\\
We investigate the structure of $S_2$ \correct{in comparison to $S_1$} and consider \correct{a set of simulated} dynamic concentrations to survey whether neglecting the second term in the new model may be justified. Before looking at example reconstructions to compare reconstructions with the old and extended model in Sec.\ref{sec:examples}, we are looking at the influence of the new summand in the dynamic model.
%
%
\begin{table}[htbp]
	\caption{Physical parameters used for the simulations}
	\label{tab:params}
	\begin{tabular}{llll}
		\hline\noalign{\smallskip}
		\textbf{Parameter} & &\textbf{Value} &\textbf{cf.}\\ 
		\noalign{\smallskip}\hline\noalign{\smallskip}
		\textit{Constants} & & \\ 
		\noalign{\smallskip}\hline\noalign{\smallskip}
		Permeability constant &$\mu_0$ &$4\pi\cdot10^{-7}$ N/A$^2$\\
		Boltzmann constant   &$k_B$   &$1.38064852\cdot10^{-23}$ J/K \\ 
		\noalign{\smallskip}\hline\noalign{\smallskip}
		\textit{Particles} & & \\ 
		\noalign{\smallskip}\hline\noalign{\smallskip}
		Temperature   & $T$   &$310$ K &\cite{knopp2009model}\\
		Saturation magnetization   &$M_C$   &$\frac{0.6}{\mu_0}$ T &\cite{knopp2009model}\\
		Particle core diameter   &$D$   &$20\cdot10^{-9}$ m &\cite{eberbeck2012multicore}\\
		Particle core volume   &$V_C$   &$\frac{1}{6}\pi D^3$ m$^3$ &\cite{knopp2009model}\\
		Particle magnetic moment   &$\alpha$   &$M_C V_C$ Am$^2$ &\cite{knopp2009model}\\
		Parameter of Langevin function  &$\beta$   &$(k_B T)^{-1}$ N$^{-1}$m$^{-1}$  &\cite{knopp2009model}\\ 
		\noalign{\smallskip}\hline\noalign{\smallskip}
		\textit{Scanner} & & &\cite{knoppOpenMPI}\\ 
		\noalign{\smallskip}\hline\noalign{\smallskip}
		Excitation frequencies   &$[f_x,f_y,f_z]$   &$[2.5/102,  2.5/96,  2.5/99]$ MHz\\
		Excitation amplitudes   &$[a_x,a_y,a_z]$   & $[0.012,0.012,0.0]$ T\\
		Excitation phase shifts   &$[\varphi_x,\varphi_y,\varphi_z]$   &$[\frac{\pi}{2},\frac{\pi}{2},\frac{\pi}{2}]$ \\
		Gradient strengths   &$[g_x,g_y,g_z]$   &$[-1,-1,2]$ T/m \\
		Excitation repetition time   &$T_c$   & $652.8\cdot10^{-6}$ s\\
	\end{tabular}
\end{table}\\
For the simulation, the system functions with $19\times19\times 1$ voxels and $1632$ sampling points in time are modeled according to the Langevin model \correct{using the parameters listed in Tab.\ref{tab:params}.}
%
%
\paragraph{\correct{Langevin model}}
In the Langevin model the particles are assumed to be in thermal equilibrium and the applied magnetic fields to be static. The mean magnetic moment at spatial position $r$ and time point $t$ is given by
\begin{align*}
	\bar{m}(r,t)&=\mathcal{L}_{\alpha,\beta}(\|H(r,t)\|_2)\frac{H(r,t)}{\|H(r,t)\|_2}, \qquad \bar m\in H^1(\Omega\times\mathbb{R}^+),\,\mathcal{L}_{\alpha,\beta}\in H^1(\mathbb{R}),
\end{align*} 
with $\mathcal{L}_{\alpha,\beta}:\mathbb{R}\mapsto\mathbb{R}$ being the Langevin function
\begin{align*}
	\mathcal{L}_{\alpha,\beta}(z)&=\begin{cases}
		\alpha \coth(\alpha\beta z)-\frac{1}{\beta z}, &\text{if } z \neq 0\\
		0, &\text{if } z=0
	\end{cases}
\end{align*}
with $\alpha,\beta\in \mathbb{R}$ being particle dependent parameters.\\
Eq.(\ref{eqn:newModelFD}) shows a sum of two convolutions. \correct{In a first step}, we are interested in the shape of the convolution kernels. Therefore, we compute
\begin{align*}
	\max_{r \in \Omega}\{|\hat{S}_l(r,k)|\}, \qquad l\in \{1,2\}
\end{align*} 
which are shown in Fig.\ref{fig:SMswithcurves} together with an approximation of their convex hulls. The approximation of the convex hull was calculated by determining and connecting the maximum values within the next 15 frequency steps to include all peaks of the function. Both matrices exhibit a similar structure. The peaks have the same distances ($\approx 15$ frequency steps) and the convex hull (orange line) has a similar shape. The full-width-at-half-maximum (FWHM) of the convex hull is the same ($\approx 33$ frequency steps) while the maximum of the second system matrix $\hat S_2$ is $10^4$ smaller than the maximum of the first system matrix $\hat S_1$. Thus, on a first view, the assumption that the second term $S_2$ in the dynamic model is negligible might be reasonable.
\\
In a second step, we looked at four types of dynamic concentrations during one cycle for a single voxel. The plots in Fig.\ref{fig:concAmp} show the concentration over the scan time, its time derivative and the respective Fourier transforms for each example concentration.\\
Example concentration 1 is depicted in Fig.\ref{fig:concShort} which shows one peak at the beginning of the scan. The tracer is flowing through the voxel for a short period of time. This could be a small tracer bolus moving fast through the volume of the voxel. In the second example, shown in Fig.\ref{fig:concLong}, the concentration increases strongly in the beginning, remains constant for a short period of time and decreases again. The tracer flows through the voxel for a longer period of time. Example 2 represents a larger bolus moving fast through the voxel. Example 3, shown in Fig.\ref{fig:concSlow}, shows a slow increase and decrease of the concentration. This represents a slowly moving small bolus. Example 4 is a periodic version of the first example. Fig.\ref{fig:concShortPeriodic} shows two peaks within the scan time. The tracer flows two times through the voxel with a high velocity. This represents a small bolus with fast periodic motion.\\
Looking at the Fourier transformations shows that the maximal absolute values of the Fourier transformed concentrations $\hat c$ are about $10^4$ smaller than the maximal absolute values of the Fourier transformed time derivatives $\frac{\widehat{dc}}{dt}$ for all 4 examples. This demonstrates that for these dynamic concentrations the magnitude of the two summands of the new dynamic model is the same.\\
The imaging process in Fourier space is a convolution of the Fourier transformed system matrices with the Fourier transformed concentration and its time derivative. Thus, the concentration is smoothed by the system matrix. The kernels $S_1$ and $S_2$ have the same width meaning that the concentration and its derivative are smoothed equally.
\begin{figure}[htbp]
	\centering
	\subfloat[Example 1: The tracer flows through the voxel for a short period of time\label{fig:concShort}]{
		\includegraphics[width=.9\textwidth]{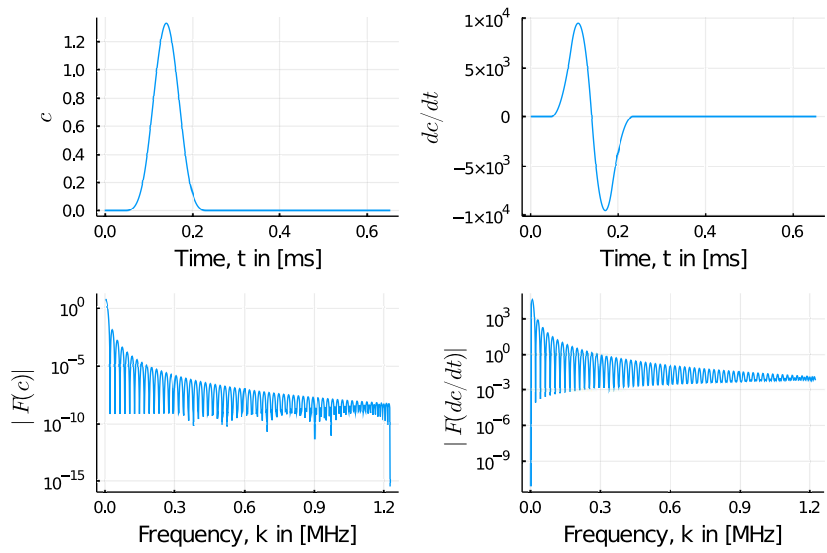}}\\
	\subfloat[Example 2: The tracer flows through the voxel for a longer period of time\label{fig:concLong}]{
		\includegraphics[width=.9\textwidth]{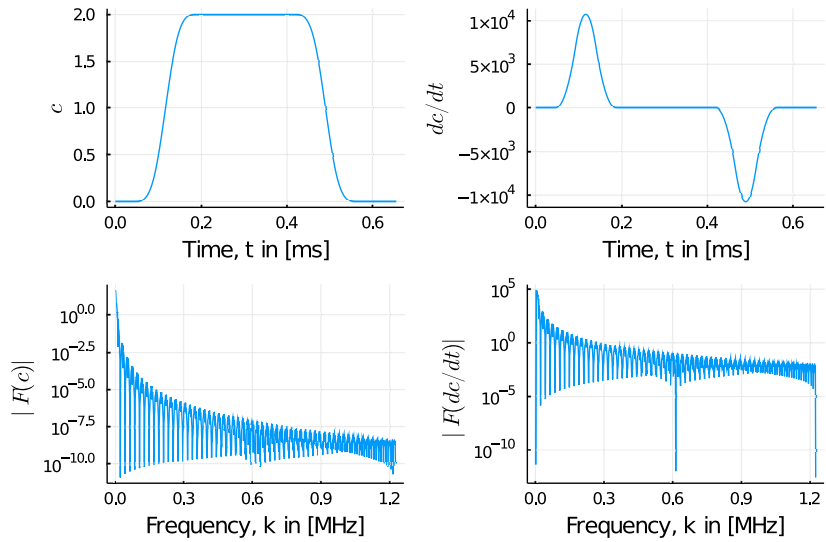}}\\
	\caption{Example concentrations and their time derivatives in time and frequency domain. The dynamics appear within one scanning cycle.}
\end{figure}
\begin{figure}[htbp]\ContinuedFloat
	\centering
	\subfloat[Example 3: The tracer accumulates and dissipates slowly in the voxel\label{fig:concSlow}]{
		\includegraphics[width=.9\textwidth]{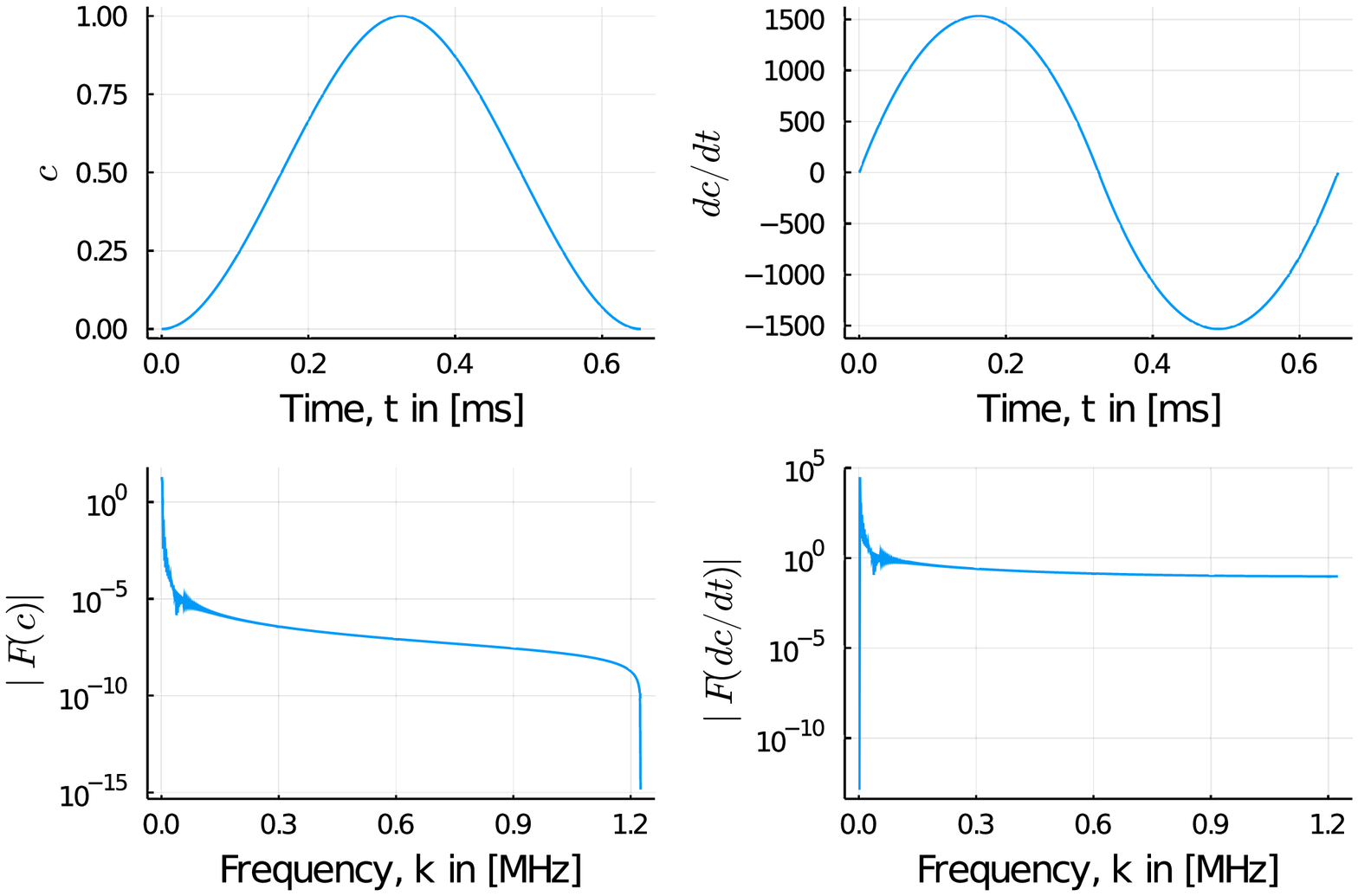}}\\
	\subfloat[Example 4: The tracer flows through the voxel for a short period of time for two times\label{fig:concShortPeriodic}]{
		\includegraphics[width=.9\textwidth]{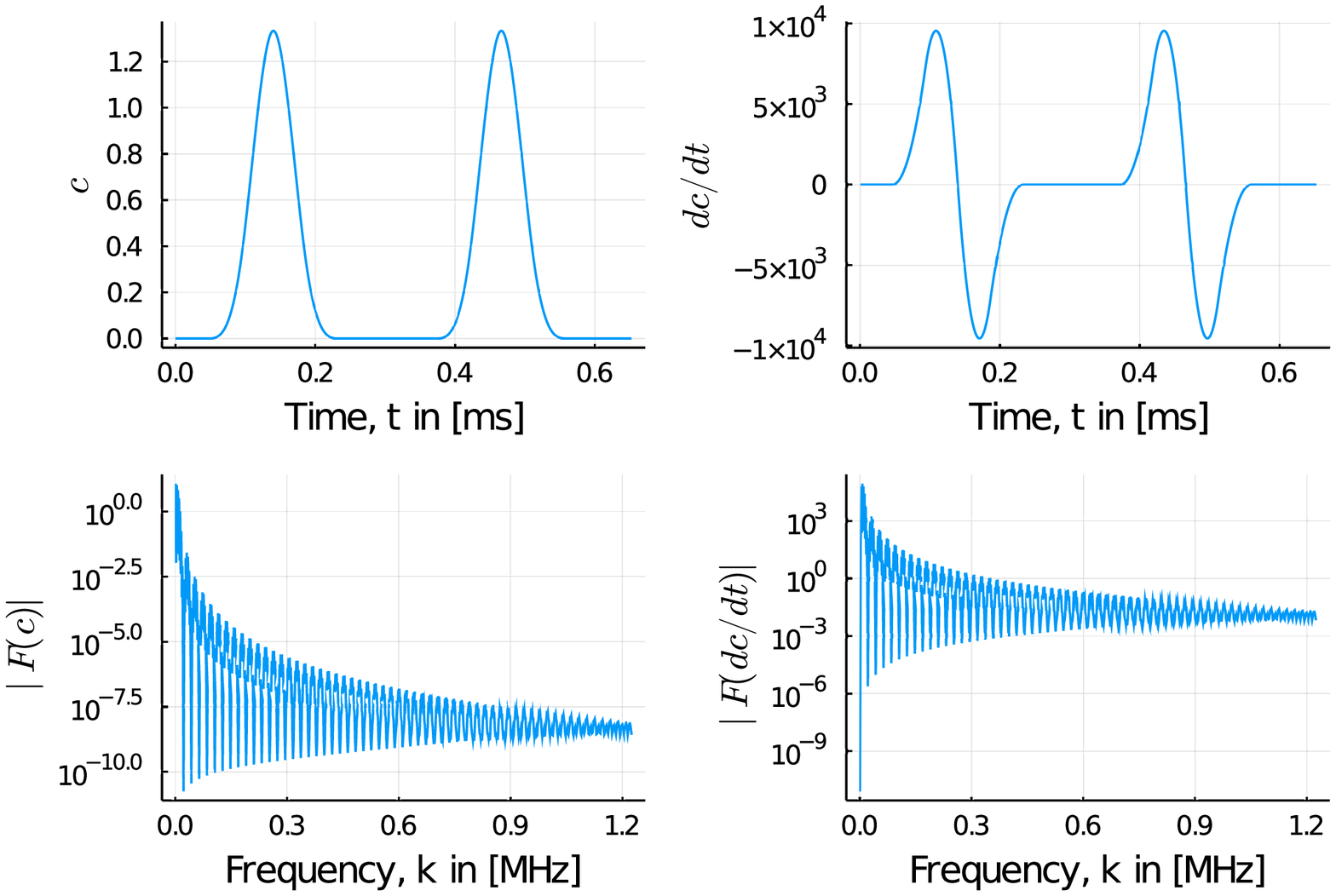}}
	\caption{Example concentrations and their time derivatives in time and frequency domain. The dynamics appear within one scanning cycle.}\label{fig:concAmp}
\end{figure}\\
%
%
%
%
%
To further examine these effects, we split up the discrete forward model such that the signal 
\begin{align*}
\hat u(k_j)&=\eta \sum_{i=1}^R \underbrace{\hat S_1(r_i,k_j)\ast\hat c(r_i,k_j)}_{=a(i,j)}+ \eta\sum_{i=1}^R \underbrace{\hat S_2(r_i,k_j)\ast\widehat{\pd{c}{t}}(r_i,k_j)}_{=b(i,j)}\\
&=\eta \underbrace{\sum_{i=1}^R a(i,j)}_A+ \eta \underbrace{\sum_{i=1}^R b(i,j)}_B, \qquad j=1,\dots,n_K,
\end{align*}
is now the sum of $A$ and $B$, where $A$ denotes the signal component generated by the first system matrix $S_1$ and $B$ the signal component generated by the second system matrix $S_2$. The convolution of the frequency components of system matrix 1 and the tracer distribution is named $a$ and the convolution of system matrix 2 with the derivative of the concentration is named $b$.
%
%
\begin{figure}[htbp]
	\centering
	\subfloat[$a=\hat S_1(r_i,k_j)\ast\hat c(r_i,k_j)$\label{fig:compareConvolutionsA}]{
		\includegraphics[width=.41\textwidth]{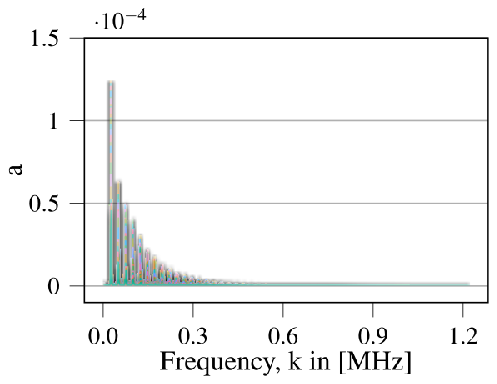}}
	\qquad
	\subfloat[$b=\hat S_2(r_i,k_j)\ast\widehat{\pd{c}{t}}(r_i,k_j)$\label{fig:compareConvolutionsB}]{	
		\includegraphics[width=.41\textwidth]{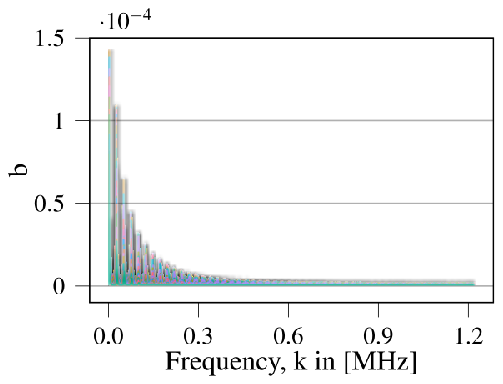}}
	\caption{Absolute values of the convolution of the system matrices with the concentration and its derivative. Each curve shows the frequency amplitudes for one of the $19^2$ voxels.}\label{fig:compareConvolutions}
\end{figure}\\
Using the dynamic example concentration 3 shown in Fig.\ref{fig:concSlow}, Fig.\ref{fig:compareConvolutions} shows $a$ and $b$ for each voxel. As expected, one can see that the shape and the maximum values of both terms are similar. Both plots show maximum values of about $1.2\cdot10^{-4}$.
%
%
\begin{figure}[htbp]
	\centering
	\subfloat[Signal part $A=\sum_{i=1}^R \hat S_1(r_i,k_j)\ast\hat c(r_i,k_j)$ generated by the first system matrix\label{fig:compareSumA}]{
		\includegraphics[width=.39\textwidth]{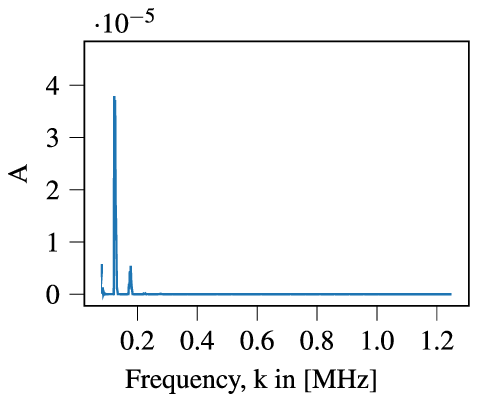}}
	\qquad
	\subfloat[Signal part $B=\sum_{i=1}^R \hat S_2(r_i,k_j)\ast\widehat{\pd{c}{t}}(r_i,k_j)$ generated by the second system matrix\label{fig:compareSumB}]{
		\includegraphics[width=.39\textwidth]{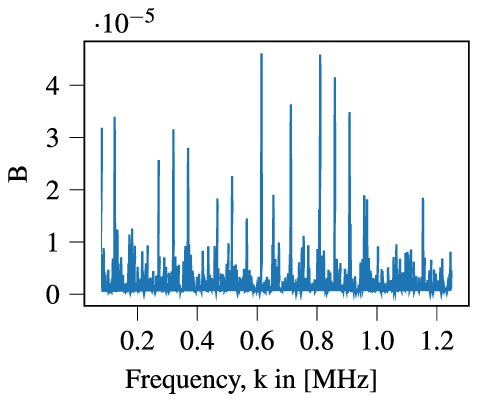}}
	\caption{Comparison of the two summands of the dynamic forward model $\hat u(k_j)=\eta(A+B)$. A frequency selection with $k\in[0.08,1.25]$MHz is shown.}\label{fig:compareAmplitudesOfSMterms}
\end{figure}
Fig.\ref{fig:compareAmplitudesOfSMterms} shows the plots of $A$ and $B$ for the frequencies $k\in[0.08,1.25]$MHz. Again, both terms have the same order of magnitude. For $A$ the frequencies with high amplitudes have a small variance while for $B$ high amplitudes can be observed in the whole frequency range. This shows that even for this example with slower dynamics the second component of the forward model has a significant impact on the signal. Thus, the second summand in Eq.\eqref{eqn:newcontinuousmodel} should not be generally neglected for dynamic concentrations.

%% file: Recos.tex
\begin{figure}[htbp]
	\centering
	\includegraphics[height=.2\textwidth]{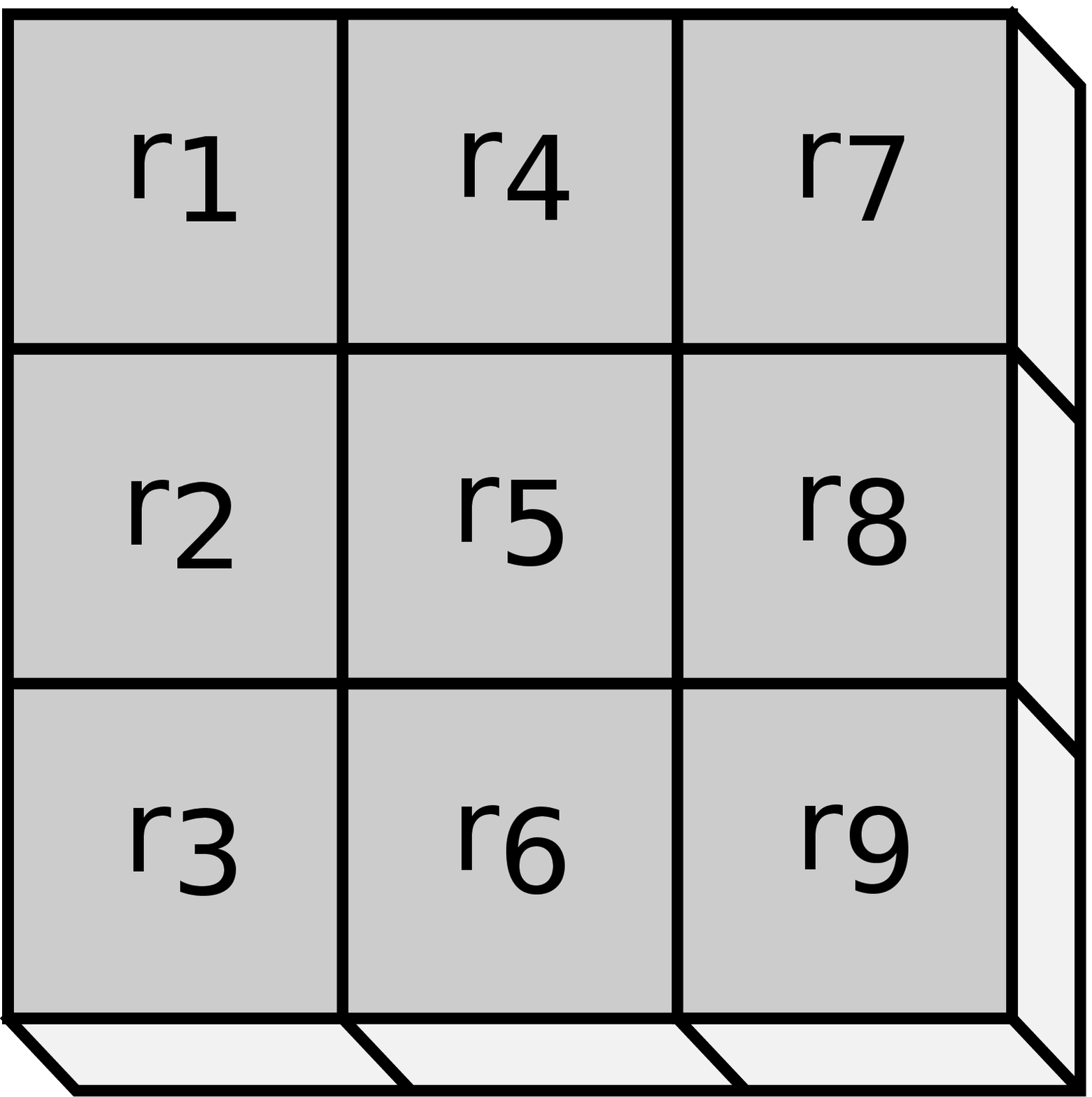}
	\caption{Spatial setup of the One- and three-peak phantom. They consist	of $3\times 3 \times 1$ voxels indexed \mbox{from 1 to 9.}}\label{fig:phantSetup}
\end{figure}

%
The challenge in solving the dynamic inverse problem (\ref{eqn:newdiscretemodel}) is the high number of degrees of freedom. We therefore use a minimalist setup with a grid of $3\times3\times1$ voxels. We use two computational phantoms to simulate an MPI measurement with the dynamic model and reconstruct it with the dynamic and the static model. They are named \textit{one-peak phantom} and \textit{three-peak phantom} and their spatial setup can be seen in Fig.\ref{fig:phantSetup}.\\ 
%
\subsection{Parameterized concentration curve}\label{sec:OnePeak}
\begin{figure}[htbp]
	\centering
	\subfloat[The tracer distribution of the central voxel~$r_5$ of one-peak phantom 1F changes during the scan of the first frame. The remaining voxels have a constant tracer concentration of zero.\label{fig:phantA}]{
		\includegraphics[width=.41\textwidth]{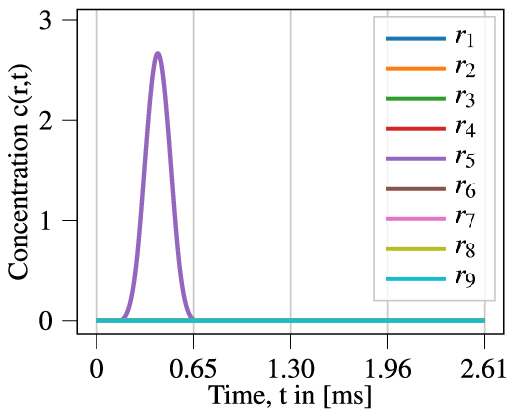}}
	\qquad
	\subfloat[The tracer distribution of the central voxel~$r_5$ of the three versions of the one-peak phantom.\label{fig:phantAvoxel5}]{
		\includegraphics[width=.41\textwidth]{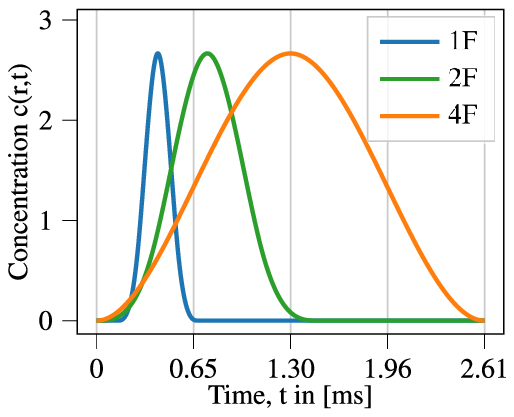}}
	\caption{The three versions of the one-peak phantom differ only in the width of the concentration peak of voxel~$r_5$, while the remaining voxels have a constant tracer concentration of zero.}\label{fig:onePeakPhant}
\end{figure}

To reduce the degrees of freedom and computational cost, the tracer concentration of the one-peak phantom is described by parametric curves 
\begin{align*}
	c(r_i,t)&=\sum_m b_{m,r_i}\psi_m(t)=\mathbf{c}(\lambda_B(r_i),t)  \qquad \in L^2(\mathbb{R}^3)\times C^2(\mathbb{R})
\end{align*}
for $i=1,\dots,R$, with parameter set $\lambda_B=[b_{m,r_i}]_{i=1,\dots,R}^{m=1,\dots,M}$, \correct{$M\in\mathbb{N}$} and basis functions $\psi_m$ which are cubic B-splines. This means that for each voxel $r_i$ there is a set of coefficients $b_m$ which together with the spline basis form a continuous concentration curve in time. Consequently, we assume that the concentration is twice differentiable with respect to $t$ which is a stronger condition than previously assumed in the dynamic model.\\
Cubic B-spline curves are well suited to model the dynamics of the magnetic tracer. In \cite{gravier2007}, the authors deal with the reconstruction of spatiotemporal tracer distributions in Single Photon Emission Computed Tomography (SPECT). Cubic B-spline curves are used to describe and reconstruct the dynamic distribution of the radioactive tracer from gated cardiac SPECT sequences.
%
%
\begin{table}[htbp]
	\caption{Reconstruction parameters}
	\label{tab:recoParams}
	\begin{tabular}{lll}
		\noalign{\smallskip}\hline\noalign{\smallskip}
		\textbf{Parameter} & & \textbf{Value}\\
		\noalign{\smallskip}\hline\noalign{\smallskip}
		Voxel size (phantoms and system matrices) &  &$0.0107\times0.0107\times0.0107$ m$^3$\\
		Field of view & & $[0.0320, 0.0320, 0.0107]$ m\\
		Time sampling per cycle& $n_T$ & 408\\
		Transition time between frames & $\Delta_f$& 0 s\\
		Number of frames Sec.\ref{sec:OnePeak}& F & 4\\
		Number of frames Sec.\ref{sec:ThreePeak}& F & 10\\
	\end{tabular}
\end{table}
\correct{We generate three variants of the one-peak phantom 1F, 2F and 4F. For  one-peak phantom 1F the coefficients for all $R=9$ voxels except $r_5$ are zero. The concentration is non-zero within the scan time of one frame. Fig.\ref{fig:phantA} shows the development of the tracer distribution for the total scan time where each curve describes the concentration within one voxel. The plot shows a peak at $t=0.4128$ms with a concentration of $2.67$ for voxel $r_5$. Versions 2F and 4F differ only in the width of the concentration peak. The concentration peak for $r_5$ lasts for the scan time of 2 frames in version 2F and 4 frames in version 4F (see Fig.\ref{fig:phantAvoxel5}). The three variants can be related to boluses with different velocities. The bolus in version 1F is twice as fast as in 2F and four times faster than in version 4F.}
%
Measurements with 4 frames which are each sampled at 408 time points and the dynamic forward model with $S_1 = \pd{\bar m}{t}$ and $S_2=\bar m$ are simulated according to (\ref{eqn:newdiscretemodel}). The parameters used for \correct{the simulations are listed in Tab.\ref{tab:params} and Tab.\ref{tab:recoParams}}. There is no transition time in between the frames.
%
\begin{figure}[htbp]
	\centering
	\subfloat[Reconstruction of 1F with the dynamic model using $S_1$ and $S_2$.\label{fig:onePeak1FrecoS1S2}]{
		\includegraphics[width=.41\textwidth]{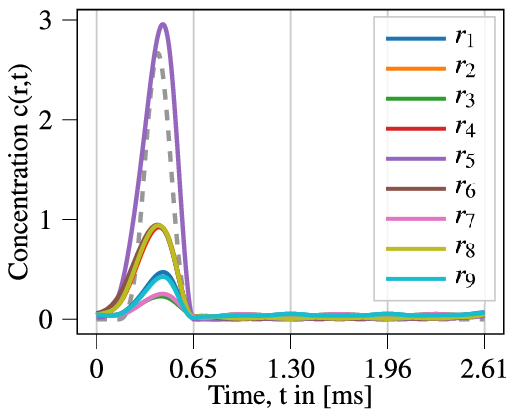}
	}
	\qquad
	\subfloat[Reconstruction of 1F  using only $S_1$.\label{fig:onePeak1FrecoS1}]{
		\includegraphics[width=.41\textwidth]{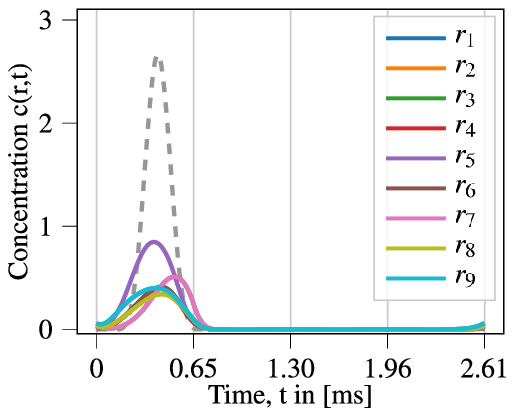}
	}\\
	\subfloat[Reconstruction of 2F with the dynamic model using $S_1$ and $S_2$.\label{fig:onePeak2FrecoS1S2}]{
		\includegraphics[width=.41\textwidth]{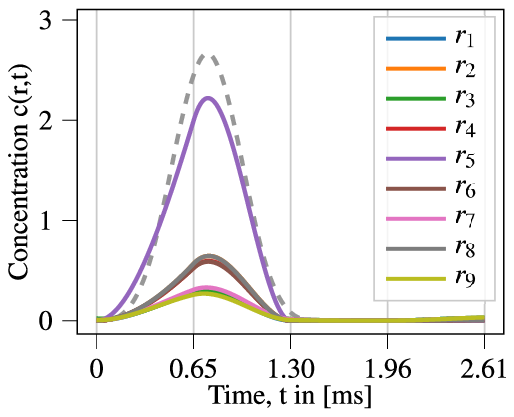}
	}
	\qquad
	\subfloat[Reconstruction of 2F  using only $S_1$.\label{fig:onePeak2FrecoS1}]{
		\includegraphics[width=.41\textwidth]{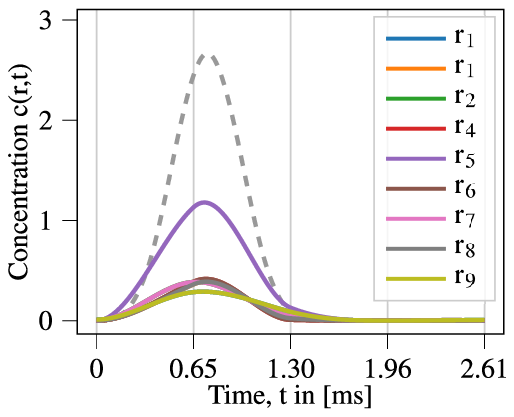}
	}\\
	\subfloat[Reconstruction of 4F with the dynamic model using $S_1$ and $S_2$.\label{fig:onePeak4FrecoS1S2}]{
		\includegraphics[width=.41\textwidth]{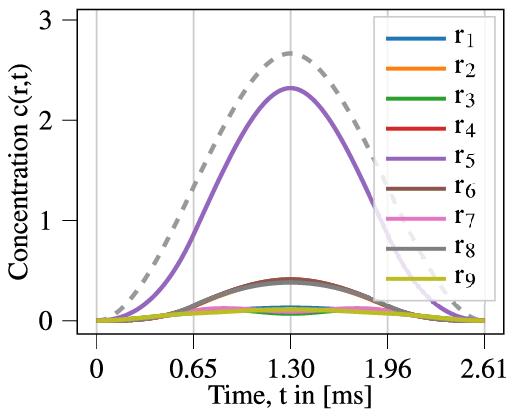}
	}
	\qquad
	\subfloat[Reconstruction of 4F using only $S_1$.\label{fig:onePeak4FrecoS1}]{
		\includegraphics[width=.41\textwidth]{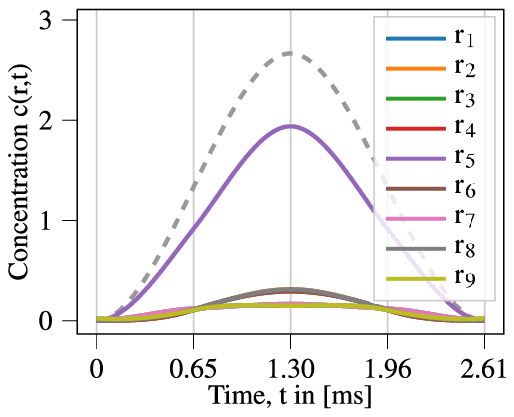}
	}
	\caption{Measurements of the dynamic one-peak phantoms 1F, 2F and 4F are simulated with the dynamic forward model (\ref{eqn:newdiscretemodel}). They are reconstructed with either both $S_1$ and $S_2$ (left) or only $S_1$ (right). All plots show averages of x- and y-channel reconstructions. The dashed lines outline the true concentration in voxel~$r_5$ and the vertical grid lines mark the start and end of frames.}\label{fig:parametricRecos}
\end{figure}\\
We reconstruct the concentrations by minimization with respect to the parameter set $\lambda_B$ so that we get continuous concentration curves. The solution set is restricted to parametric spline curves in $L^2(\mathbb{R}^3)\times C^2(\mathbb{R})$ which is an implicit regularization. The curves are reconstructed with two different settings. In the first experiment both matrices are used for reconstruction which corresponds to minimizing
\begin{align*}
	\min_{\lambda_B}&\biggl\|\Bigl[\sum_{i=1}^R  S_1(r_i,t_j)\mathbf{c}(\lambda_B(r_i),t_j)+S_2(r_i,t_j)\pd{\mathbf{c}}{t}(\lambda_B(r_i),t_j)\Bigr]_{j=1,\dots,n_T}- \mathbf{u} \biggr\|_2^2
\end{align*}
with $\mathbf{u}\in \mathbb{R}^{n_T}$. The problem is minimized with 200 iterations of a conjugate gradient algorithm and no further regularization. Fig.\ref{fig:onePeak1FrecoS1S2} shows the average of reconstructions of the x- and y-channel of one-peak phantom 1F. The peak for voxel $r_5$ is located at $t=0.4448$ms with a concentration of $2.96$ which is very close to ground truth. In the same period of time also the concentration of the remaining voxels is non-zero. The peaks of the voxels with even indices have concentration values of about $0.9$ and the peaks of the voxels with odd indices have even smaller values of about $0.4$. Even if these voxels have a non-zero concentration, it is significantly lower than the value of $r_5$, so that we can expect sufficient contrast in the reconstructed images. The values for the off-diagonal voxels (voxels with even indices, cf. Fig.\ref{fig:phantSetup}) show higher concentration values than the ones on the diagonal. The x-channel reconstruction locates the concentration correctly in x and the y-channel reconstruction locates the concentration correctly in y. Thus, the off-diagonal voxels are masked by the high concentration in the central voxel.\\
In the next experiment the same measurement is reconstructed only with $S_1$ which corresponds to  minimizing
\begin{align*}
	\min_{\lambda_B}&\biggl\|\Bigl[\sum_{i=1}^R  S_1(r_i,t_j)\mathbf{c}(\lambda_B(r_i),t_j)\Bigr]_{j=1,\dots,n_T}- \mathbf{u} \biggr\|_2^2
\end{align*}
with $\mathbf{u}\in \mathbb{R}^{n_T}$. The problem is again minimized with 200 iterations of a conjugate gradient algorithm and no further regularization. The result for the average of x- and y-channel reconstructions is shown in Fig.\ref{fig:onePeak1FrecoS1}. The reconstruction shows a peak for voxel $r_5$ at $t=0.3808$ms which is close to the ground truth but with a significantly smaller concentration of $0.85$. Again there are concentration peaks for all remaining voxels with values of about $0.4$. This means that the reconstructed images will show reduced contrast. And the true concentration is underestimated.
%
\begin{figure}[htbp]
	\centering	
	\subfloat[Phantom\label{fig:phantImage1F}]{
		\includegraphics[height=.25\textwidth]{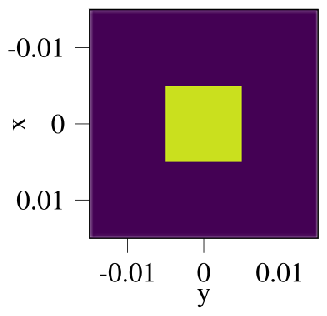}}
	\qquad
	\subfloat[Reconstruction using $S_1$~and~$S_2$\label{fig:recoImage1FS1S2}]{
		\includegraphics[height=.25\textwidth]{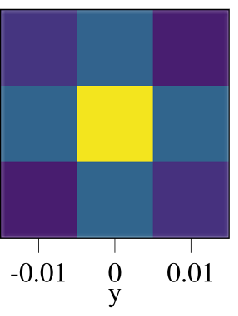}}
	\qquad
	\subfloat[Reconstruction using only~$S_1$\label{fig:recoImage1FS1}]{
		\includegraphics[height=.25\textwidth]{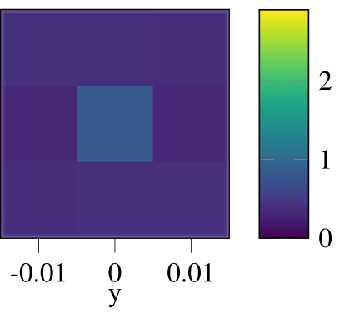}}
	\caption{Reconstructions of the one-peak phantom 1F at the time point of the concentration peak $t=0.4128$ms}\label{fig:recoOnePeakF1}
\end{figure}\\
To get a more intuitive impression of the impact of the discussed curves on the reconstruction quality, Fig.\ref{fig:recoOnePeakF1} shows a frame of the phantom and the two reconstructed time-series at the time point of the maximum concentration ($t=0.4128$ms). Looking at the image of the first experiment in Fig.\ref{fig:recoImage1FS1S2} one can observe a good contrast and slightly higher concentration values for the off-diagonal voxels. As discussed above, the image from the second experiment shown in Fig.\ref{fig:recoImage1FS1} exhibits poor contrast and significantly lower concentration values compared to the phantom.
%
\begin{figure}[htbp]
	\centering	
	\subfloat[Phantom\label{fig:phantImage4F}]{
		\includegraphics[height=.25\textwidth]{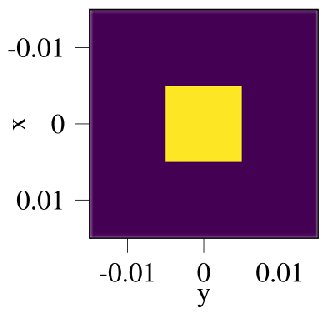}}
	\qquad
	\subfloat[Reconstruction using $S_1$~and~$S_2$\label{fig:recoImage4FS1S2}]{
		\includegraphics[height=.25\textwidth]{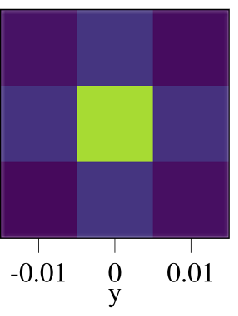}}
	\qquad
	\subfloat[Reconstruction using only~$S_1$\label{fig:recoImage4FS1}]{
		\includegraphics[height=.25\textwidth]{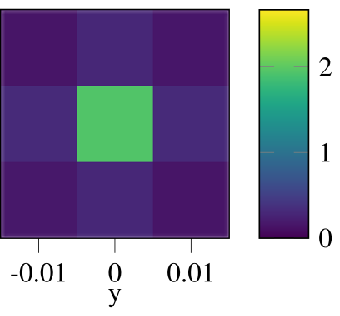}}
	\caption{Reconstructions of one-peak phantom 4F at the time point of the concentration peak $t=1.304$ms}\label{fig:recoOnePeakF4}
\end{figure}\\
%
\correct{Version 2F and 4F of the one-peak phantom are reconstructed analogously to version 1F. While also for version 2F reconstructed with both matrices the concentration peaks show the correct location and $83\%$ of the true amplitude (see Fig.\ref{fig:onePeak2FrecoS1S2}), the peak in the reconstruction using only $S_1$ is less than $50\%$ of ground truth (see Fig.\ref{fig:onePeak2FrecoS1}). The reconstructions for version 4F with and without $S_2$ are shown in Fig.{\ref{fig:onePeak4FrecoS1S2}} and \ref{fig:onePeak4FrecoS1}. The location is correct for both reconstruction methods and the amplitude of the peak for $r_5$ reaches $87\%$ of the ground truth for reconstruction with both matrices and $73\%$ for using only $S_1$. Also the concentrations for the remaining voxels are sufficiently low in both cases. Looking at a frame of one-peak phantom 4F and the two reconstructed time-series at the time point of the maximum concentration ($t=1.304$ms) in Fig.\ref{fig:recoOnePeakF4}, one can see that the contrast in the reconstruction without $S_2$ (see. Fig.\ref{fig:recoImage4FS1}) is improved compared to version 1F and almost comparable to the reconstruction using $S_1$ and $S_2$ (see. Fig.\ref{fig:recoImage4FS1S2}).
%
\begin{figure}[htbp]
	\centering
	\includegraphics[height=.1\textwidth]{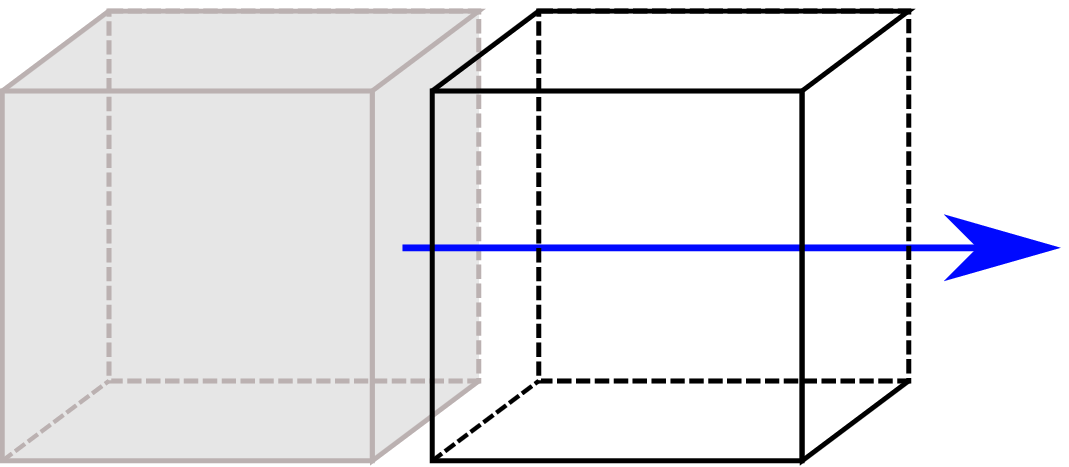}
	\caption{Simplified bolus moving through a voxel with constant velocity $v$.}\label{fig:bolus}
\end{figure}\\
In order to get an impression on the strength of the dynamics in one-peak phantom 4F, we relate these values to a $2\times 2\times 2$ mm$^3$ bolus with constant concentration $c_{max}$ moved through a $2\times 2 \times 2$ mm$^3$ voxel with a constant velocity $v$ (see Fig.\ref{fig:bolus}). The one-peak phantom has a maximum concentration of $c_{max}=2.667$ and a maximum time derivative of $\dot{c}_{max}=3065$. The average change rate yields a bolus velocity of $v_{av}=2\cdot10^{-3}/(2\cdot T_c)=1.53$ m/s and the maximum change rate results in a velocity of $v_{max}=2\cdot10^{-3}/(c_{max}/\dot{c}_{max})=2.3$ m/s. Thus, we state $v_{dyn}=1.53$ m/s as a preliminary threshold velocity. For reconstructions with an average flow $v>v_{dyn}$ the dynamic model will improve the reconstruction quality in comparison to the static model.}
%
\subsection{Frame-by-frame reconstruction}\label{sec:ThreePeak}
\begin{figure}[htbp]
	\subfloat[The concentration of the three-peak phantom changes in time only in voxel~$r_4$, $r_5$ and $r_6$. In the remaining voxels the concentration is zero. The time-shifted concentration peaks form a motion from $r_4$ to $r_6$.\label{fig:3peakPhant}]{
		\includegraphics[width=.4\textwidth]{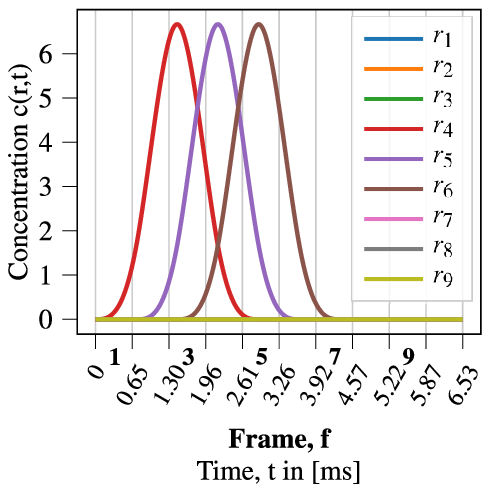}
	}\\
	\qquad
	\subfloat[Average of x- and y-channel frame-by-frame reconstructions with the dynamic model using $\mathbf{S}_1$,  $\mathbf{S}_2$\label{fig:recoD}]{
		\includegraphics[width=.4\textwidth]{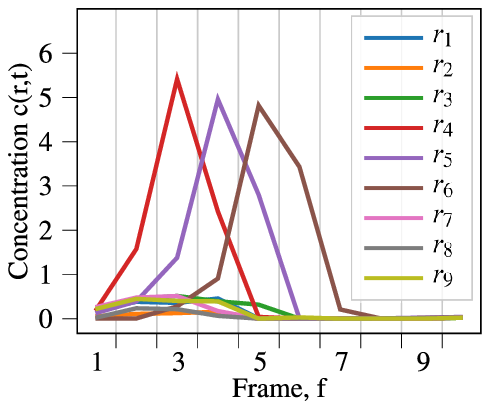}}
	\qquad
	\subfloat[Average of x- and y-channel frame-by-frame reconstructions with the static model using only $\mathbf{S}_1$\label{fig:recoA}]{
		\includegraphics[width=.4\textwidth]{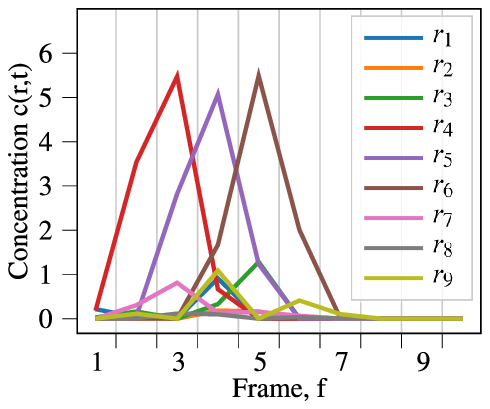}
	}
	\caption{A measurement of the dynamic three-peak phantom is simulated with the dynamic forward model (\ref{eqn:newdiscretemodel}). Each frame is reconstructed separately assuming a static tracer distribution within each frame. The frames are reconstructed with the dynamic and the static model.}\label{fig:frameByFrameRecos}
\end{figure}

Another way to investigate the impact of the new model is to reconstruct a dynamic measurement frame-by-frame with the assumption of a static concentration within each frame. Therefore we use the three-peak phantom.\\
As the one-peak phantom, the tracer concentration of the three-peak phantom is described by parametric curves $c(r_i,t)=\sum_m b_{m,r_i}\psi_m(t)=\mathbf{c}(\lambda_B(r_i),t)$ with parameter set $\lambda_B$ and basis functions $\psi_m$ which are cubic B-splines meaning that for each voxel $r_i$ there is a set of coefficients $b_m$ which together with the spline basis form a continuous concentration curve in time.\\
For the three-peak phantom only the coefficients for the voxels $r_4$, $r_5$ and $r_6$ are non-zero. The tracer distribution during the total scan time is shown in Fig.\ref{fig:3peakPhant} where each curve describes the concentration within one voxel. There is a concentration peak of $6.67$ for voxel $r_4$, $r_5$ and $r_6$. The peaks are shifted in time, such that this dynamic can be seen as an object or tracer bolus that moves from voxel $r_4$ to voxel $r_6$ considering the location of the voxels in Fig.\ref{fig:phantSetup}. The peaks are located in the scan time of frame 3, 4 and 5 \correct{and have a temporal width of about 4 frames}. The concentration of the remaining voxels is zero.\\
A measurement with $F=10$ frames which are each sampled at 408 time points and the dynamic matrix model (\ref{eqn:newdiscretemodel}) with $S_1$ and $S_2$ is simulated. \correct{The parameters used for the simulation can be found in Tab.\ref{tab:params} and Tab.\ref{tab:recoParams}}.\\
The dynamic tracer distribution is reconstructed with two different settings. The first one uses information about the tracer dynamics from the reconstructions of previous frames and the second one reconstructs each frame independently. In fact the reconstructions are piecewise constant functions over time. For better comparison the results depicted in Fig.\ref{fig:frameByFrameRecos} show linear interpolations of the static reconstructions of 10 frames.\\
In the first experiment both matrices are used for reconstruction of each frame while the time derivative $\pd{\mathbf{c}}{t}=\frac{\mathbf{c}^f-\mathbf{c}^{f-1}}{\Delta t}$ is the divided difference of the concentration vector of the current and the preceding frame. This corresponds to minimizing \correct{
\begin{align*}
	\min_{\mathbf{c}^f}\Bigl\|\Bigl(\mathbf{S}_1 \mathbf{c}^f+\mathbf{S}_2 \frac{\mathbf{c}^f-\mathbf{c}^{f-1}}{\Delta t}\Bigr)-\mathbf{u}^f\Bigr\|_2^2, \qquad f=1,\dots,F
\end{align*}}
with $\mathbf{u}^f\in \mathbb{R}^{n_T}$, $\mathbf{c}^f\in \mathbb{R}^{R}$, $\mathbf{S}_1, \mathbf{S}_2\in \mathbb{R}^{n_T \times R}$. Fig.\ref{fig:recoD} shows the average of x- and y-channel reconstructions which were reconstructed in time domain with 100 iterations of a gradient descent algorithm and no further regularization. It can be seen that the peaks are correctly located in frame 3, 4 and 5. The amplitude of the peaks is slightly lower than the ground truth and decreasing, $5.41$ for $r_4$, $4.96$ for $r_5$ and $4.82$ for $r_6$. There is a non-zero concentration for the remaining voxels in the first 5 frames of less than $0.5$. So the reconstructed images will exhibit sufficient contrast.\\
In the next experiment the same measurement is reconstructed using only $S_1$, i.e. minimizing
\begin{align*}
	\min_{\mathbf{c}^f}\bigl\|\mathbf{S}_1 \mathbf{c}^f-\mathbf{u}^f\bigr\|_2^2, \qquad f=1,\dots,F
\end{align*}
with $\mathbf{u}^f\in \mathbb{R}^{n_T}$, $\mathbf{c}^f\in \mathbb{R}^{R}$, $\mathbf{S}_1\in \mathbb{R}^{n_T \times R}$ in time domain with 100 iterations of a gradient descent algorithm and no further regularization. The result is shown in Fig.\ref{fig:recoA}. 
Again the peaks are located correctly in frame 3, 4 and 5. The amplitudes $5.47$ for $r_4$, $5.07$ for $r_5$ and $5.49$ for $r_6$ are also slightly lower than in the phantom and differ less than in the first experiment. The remaining voxels show non-zero concentrations up to $1.2$ being more than twice as high as for the first experiment.

%% file: Conclusion.tex
%
We introduced a new extended forward model for dynamic magnetic particle imaging. It was shown that the standard forward model does not account for dynamic tracer distributions which is corrected by the extended model presented in this paper.\\
One of the main differences is that the dynamic model contains a second summand with a second system matrix. For different kinds of dynamic concentrations the two summands have been examined. The order of magnitude of the summands is the same for the chosen dynamic examples. This emphasizes the importance of the new model for dynamic tracer distributions.\\
\correct{Furthermore, we simulated measurements from dynamic concentrations with the extended model and reconstructed them with both the dynamic and the static model. In the experiments in Sec.\ref{sec:OnePeak} three simple phantoms with different change rates are examined. For one-peak 4F, the phantom with the lowest change rates, the static approach using only one system matrix provided an acceptable reconstruction quality. For the phantoms with higher  change rates, one-peak phantom 2F and 1F, the static approach resulted in reconstructions with low contrast and significantly lower amplitudes than ground truth while the dynamic approach performed well on all three phantoms. While a quantitative study of this is beyond the scope of this article, we can state that for higher change rates than in one-peak phantom 4F the dynamic model should be considered for reconstruction.}\\ 
%
The presented dynamic model is more general than existing approaches for dynamic concentration reconstruction as it is not limited to periodic motion and can be applied to motions with high velocities. While in this paper the equilibrium model is used, the dynamic model allows to incorporate more advanced magnetization models which could improve the reconstruction quality in the future. The reconstruction approach with parametric concentration curves features an implicit dynamic regularization. Additional spatial or temporal regularization, like sparsity in time and space, can be included easily.
Moreover, the model allows for joint reconstructions of the particle concentration and its time-derivative which might be of special interest for blood-flow diagnostics.\\
%
It remains future research to develop new reconstruction techniques for dynamic tracer distributions based on this model and extend it to multi-patch imaging sequences for larger volumes. Furthermore, the methods need to be evaluated for simulations with phantoms of realistic size and physical phantom measurements. Practice-oriented scenarios might require improved minimization schemes for reconstruction. \correct{The dynamic model might also be combined with measurement-based reconstruction. While using a calibration scan for $S_1$, the second system matrix can be modeled and corrected with the transfer function. Alternatively, $S_2$ might be learned from its measured time-derivative $S_1$. In addition to our preliminary statement, a quantitative study of phantoms with different velocities is required for a more precise proposition about when the dynamic model is necessary based on the level of dynamics and the desired reconstruction quality.} A further theoretical research direction is the analysis of features like the ill-posedness of the dynamic reconstruction problem.
%
%
%
%